\documentclass{iaus}
\usepackage{graphicx}

\usepackage[authoryear]{natbib}

\makeatletter
\newcommand{\fig}[1]{Fig.~\ref{#1}} \font\tenbg=cmmib10 at 10pt
\def \rvecmu{{\hbox{\tenbg\char'026}}}

\providecommand{\boldsymbol}[1]{\mbox{\boldmath $#1$}}

\newcommand{\aap}{A\&A}

\newcommand{\apjs}{ApJS}

\makeatother
\title[MHD Simulations] %% give here short title %%
{MHD simulations of disk-star interaction}

\author[Marina M. Romanova]   %% give here short author list %%
{Marina M. Romanova$^1$, M. Long$^1$, A.K. Kulkarni$^1$, R. Kurosawa
$^2$\break\
 G.V. Ustyugova$^3$, A.K. Koldoba$^3$ and R.V.E. Lovelace $^1$}

\affiliation{$^1$Astronomy Department, Cornell University,
Ithaca, NY 14853, USA; \break email: romanova@astro.cornell.edu\\[\affilskip]
$^2$Department of Physics and Astronomy, University of Nevada Las
Vegas,  Box 454002, \break 4505 Maryland Pkwy,  Las Vegas, NV
89154-4002, USA;  email: rk@physics.unlv.edu\\[\affilskip]
$^3$Keldysh Institute of Applied Mathematics, Miusskaya sq. 4,
Moscow, 125047, Russia, \break email: ustyugg@rambler.ru}

\pubyear{2007}
\volume{IAU Symposium 243}  %% insert here IAU Symposium No.
\pagerange{1--12}
\date{30 June 2007 and in revised form ??}
 \jname{Star-disk Interaction in Young Stars}
\editors{J. Bouvier \& I. Appenzeller , eds.}
\begin{document}

\maketitle

\begin{abstract}
We discuss a number of topics relevant to disk-magnetosphere
interaction and how numerical simulations illuminate them. The
topics include: (1) disk-magnetosphere interaction and the problem
of disk-locking; (2) the wind problem; (3) structure of the
magnetospheric flow, hot spots at the star's surface, and the inner
disk region; (4) modeling of spectra from 3D funnel streams; (5)
accretion to a star with a complex magnetic field; (6) accretion
through 3D instabilities; (7) magnetospheric gap and survival of
protoplanets. Results of both 2D and 3D simulations are discussed.

%% add here a maximum of 10 keywords, to be taken form the file <Keywords.txt>
\end{abstract}

\firstsection % if your document starts with a section,
              % remove some space above using this command.
\section{Introduction}

Disk accretion to a rotating star with a dipole magnetic field has
been investigated theoretically by a number of authors (e.g., Ghosh
and Lamb 1978; Camenzind 1990; K\"{o}nigl 1991; Ostriker and Shu
1995; Lovelace, Romanova \& Bisnovatyi-Kogan 1995). Recently a
number of numerical simulations have been performed assuming
axisymmetry (e.g., Hayashi, Shibata \& Matsumoto 1996; Miller \&
Stone 1997; Goodson et al. 1997; Fendt \& Elstner 2000; Matt et al.
2002; von Rekowski \& Brandenburg 2003; Yelenina, Ustyugova \&
Koldoba 2006). In some of them, accretion through funnel streams has
been clearly observed and investigated (e.g., Romanova et al. 2002;
Long, Romanova \& Lovelace 2005; Bessolaz et al. 2007, 2008; Zanni,
Bessolaz \& Ferreira 2007). Full three-dimensional (3D) MHD
simulations have been done by Romanova et al. (2003, 2004), for
which  a special Godunov-type MHD code based on the ``cubed sphere"
grid has been developed (Koldoba et al. 2002). Longer 3D simulations
have been done by Kulkarni \& Romanova (2005) and for accretion to a
star with a non-dipole magnetic field by Long, Romanova \& Lovelace
(2007, 2008). Accretion through 3D instabilities has been recently
observed in 3D simulations (Romanova, Kulkarni \& Lovelace 2007;
Kulkarni \& Romanova 2008). Spectral lines from 3D funnel streams
were calculated by Kurosawa, Romanova \& Harries (2008) where a 3D
radiative transfer grid has been projected onto the 3D MHD grid. A
magnetospheric gap may stop the inward migration of protoplanets at
the CTTSs stage (Lin, Bodenheimer \& Richardson 1996), unless
different processes increase the density inside the gap (Romanova \&
Lovelace 2006). We discuss these topics in greater detail in the
following sections.

\begin{figure}
\centering
\includegraphics[height=1.5in,angle=0]{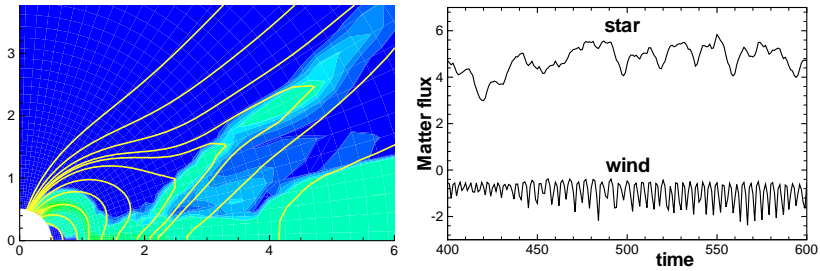}
\caption{Axisymmetric simulations of winds in the case where the
field lines are gathered in an x-type configuration. Left panel
shows the distribution of matter flux $\rho v$ and sample field
lines. Right panel shows temporal evolution of the accretion rate to
the star and to the wind, integrated over the $r=6$ surface. Time is
measured in rotational periods at $r=1$. From Romanova et al.
(2008).}\label{xwind}
\end{figure}

\section{Disk-magnetosphere interaction and disk-locking}
\label{sec:disk-mag}

\noindent{\bf Disk-magnetosphere interaction through the closed
field lines and through the funnel stream}.
 The accretion disk is
disrupted by the stellar magnetosphere and is stopped at the
distance $R_m$ from the star where the magnetic pressure is
approximately equal to the total matter pressure: $B^2/8\pi=p+\rho
v^2$ (e.g., Ghosh \& Lamb 1978; see also results of 2D and 3D
simulations, Romanova et al. 2002, 2003; Long, Romanova \& Lovelace
2005). Matter of the disrupted disk accumulates near  $R_m$ and is
lifted out of the equatorial plane into a ``funnel stream'' by the
pressure gradient force.  The matter then follows the star's dipole
field lines and is accelerated by  gravity until it hits the star's
surface (Romanova et al. 2002; 2003; Zanni et al. 2007; Bessolaz et
al. 2007). Initially, the angular momentum is carried by the funnel
stream matter, but later it is converted to angular momentum carried
by the field, so that when funnel matter arrives to the surface of
the star, almost all angular momentum is carried by magnetic field
lines and only few percent by matter (Ghosh \& Lamb 1978, 1979; see
simulations by Romanova et al. 2002; Long at el. 2005).
   Whether a star spins-up or
spins-down is determined by the ratio between the magnetospheric
radius $R_m$ and the co-rotation radius
$R_{cr}=(GM/\Omega_*^2)^{1/3}$/.
   When the star rotates slowly, and $R_{cr} >
R_m$, then the disk matter spins it up (see the main case in
Romanova et al. 2002; and  Zanni et al. 2007; Bessolaz et al. 2007).
   In this case the inner disk rotates faster than the star, and
closed magnetic field lines connecting the star and the disk form
a leading spiral which helps to spin up the star.
  All of the aforementioned
numerical simulations show a direct correlation between the
accretion rate $\dot M$ (at the surface of the star) and the spin-up
rate: the higher the accretion rate, the stronger the spin-up. This
correlation may result from the increased twist of closed magnetic
field lines connecting the inner disk and the star. If the star
rotates fast, so that $R_{cr} < R_m$, then matter is still pulled to
the funnel flow by gravity, but the star {\it spins down} (see e.g.,
Fig. 6 of Long et al. 2005 for large $\rvecmu$) because closed field
lines form a trailing spiral, so that the star loses angular
momentum. We have evidence from simulations that a significant part
of the disk-magnetosphere interaction occurs through closed field
lines connecting the star and the disk. Opening of external field
lines {\it does not} lead to disconnection of the star from the
disk.

\begin{figure}
\centering
\includegraphics[height=1.9in,angle=0]{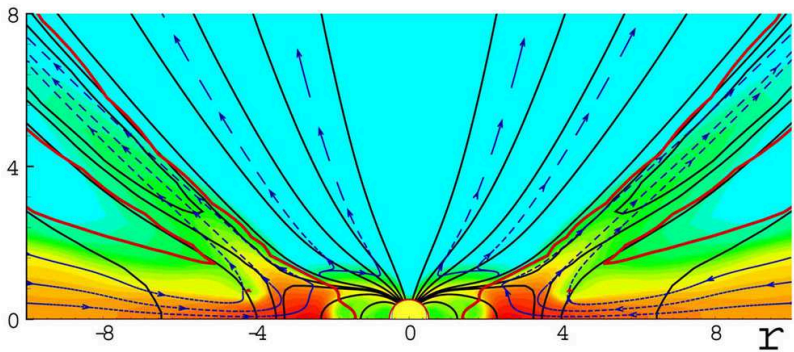}
\caption{Axisymmetric simulations of a star in the propeller regime.
The background shows the density, the thin solid lines are the
magnetic field lines, the thick solid lines show the $\beta=1$ line
(magnetic energy-density dominates in the corona and in the wind
region above the disk. The dashed lines show the streamlines of
matter flow. From Romanova et al. (2005).}\label{prop}
\end{figure}

\noindent{\bf Where does the angular momentum go?} Matter of the
disk carries significant specific angular momentum, $j_m=\Omega(R_m)
R_m^2$, which is much larger than that of the star, $j_*=\Omega(R_*)
R_*^2$, if the angular velocities are comparable, $\Omega(R_m) \sim
\Omega(R_*)$. The question is whether all the angular momentum of the
disk, which is carried initially by matter, goes to the star, or
whether a significant part of it returns to the disk via
the magnetic field, as suggested by Ostriker \& Shu (1995).
Numerical simulations have not given a final answer to this question
yet, but the last possibility appears  probable.
  In this case angular
momentum should be constantly removed from the disk by viscosity,
some type of waves propagating through the disk, or by winds. X-type
winds may be an efficient mechanism of angular momentum removal from
inner regions of the disk (Shu et al. 1994) or, winds may flow from
the disk at any distance (Pudritz \& Norman 1986; Lovelace, Romanova
\& Bisnovatyi-Kogan 1995; Lamzin et al. 2004; Ferreira, Dougados \&
Cabrit 2006). If angular momentum does not return back to the disk,
then the star will spin up and other mechanisms are required to spin
it down. This question needs further investigation.

\noindent{\bf Inflation of field lines and spinning-down}.
 Differential
rotation between the star and the disk leads to the inflation of
field lines (Aly 1980; Shu et al. 1994; Lovelace, et al. 1995). Some
field lines are strongly inflated and do not transport information
between the star and the disk. Other field lines are only partially
open and may transport angular momentum between the star and the
disk (e.g., the ``dead zone" of Ostriker and Shu 1995). Field lines
may be only partially open because, for example, the magnetic
pressure force leading to opening is balanced by the magnetic
tension force opposing the opening. In addition, inflated field
lines may reconnect and then inflate again, so that the
magnetosphere may not be stationary, but may experience
quasi-periodic or episodic restructuring (e.g. Aly \& Kuijpers 1990;
Uzdensky 2002). Strongly inflated field lines connect the star with
the slowly rotating corona, so that angular momentum always flows
out from the star to corona. The corona has a low density  and is
probably magnetically dominated, so that angular momentum may flow
from the star to the corona due to the magnetic stress, $\dot
L_f=\int{dS\cdot rB_\phi B_p/4\pi}$. A star may also lose angular
momentum due to
 major matter outflow from its surface or due to
re-direction into the outflow of some of the matter coming to the
star through the funnel flow (Matt \& Pudritz 2005) where a Weber \&
Davis (1967) type of angular momentum loss is investigated. In these
models it is important to understand the force which lifts the
matter (up to 10\% of the accretion rate) from the vicinity of the
star. Compared to the ``material" wind of Matt \& Pudritz (2005),
the angular momentum carried by the {\it magnetic stress} does not
require much matter flowing from the star and does not require dense
matter in the corona. Estimations show that only $n=10^5 1/cm^3$ are
required to support a current associated with twist of magnetic
field lines. Thus, angular momentum always flows to the ``magnetic
tower" (see also Lynden-Bell 2003) and  the torque depends on the
magnetic field strength and angular velocity of the star. In Long et
al. (1995), matter dominates in most of the corona, and spin-down
through open field lines does depend on the coronal density. However
in the propeller regime (Romanova et al. 2005; Ustyugova et al.
2006), the corona is magnetically-dominated and a significant torque
is associated with magnetic tower.

\begin{figure}
\centering \resizebox{5.9cm}{!}
{\includegraphics{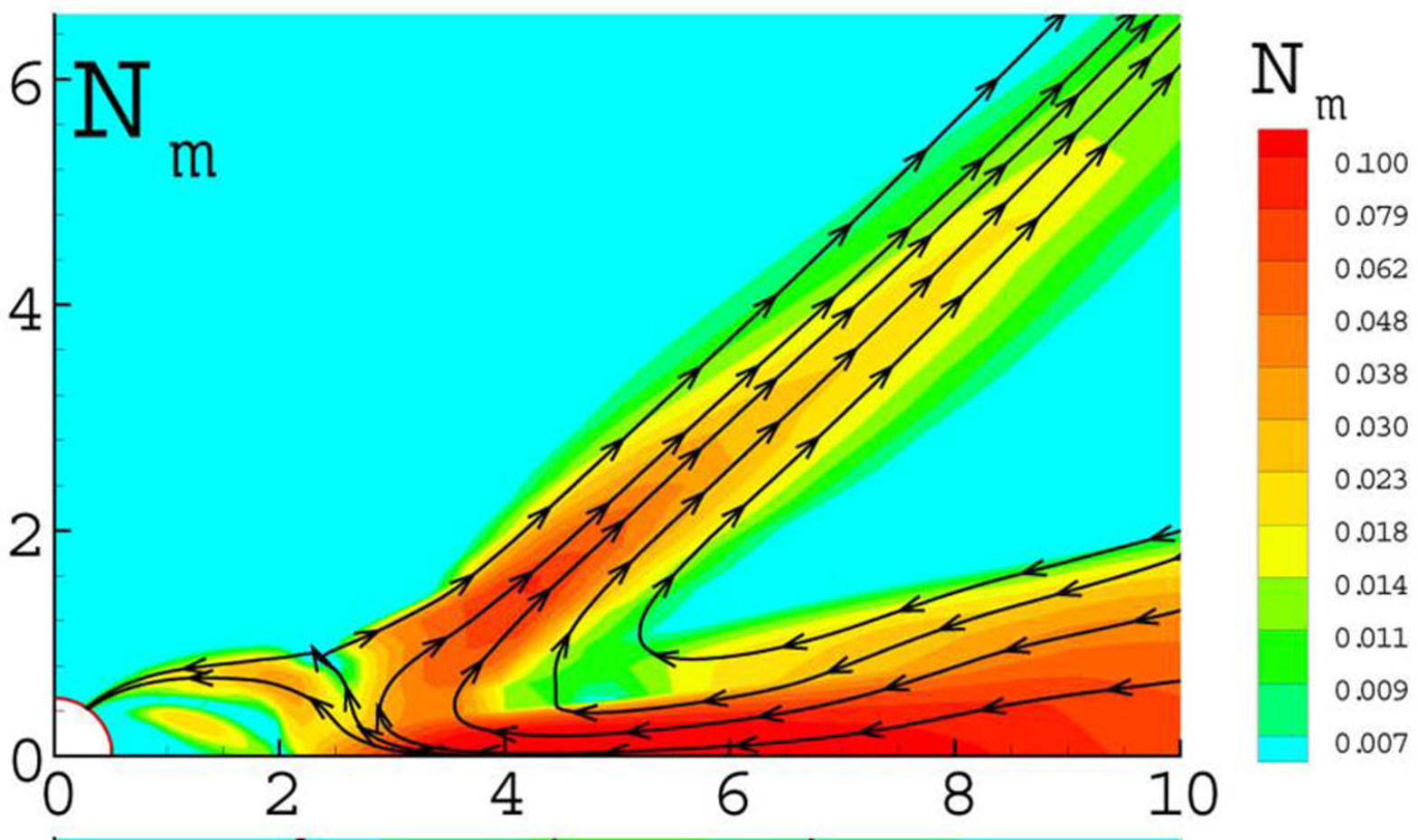} }
\resizebox{6.5cm}{!}{\includegraphics{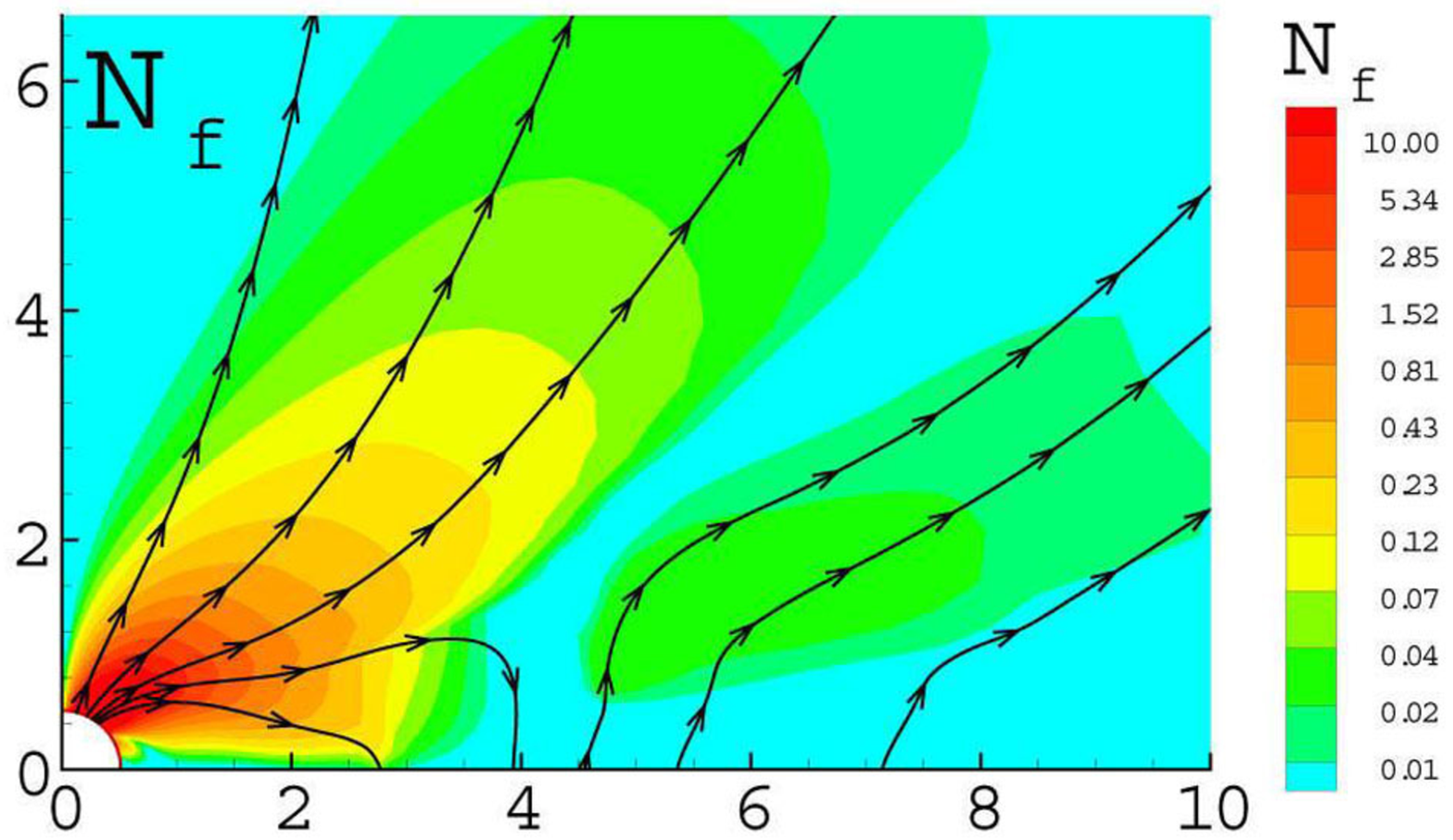} }
\caption[]{The background shows torques associated with matter $N_m$
and the magnetic field $N_f$ in the propeller regime. The
streamlines show the direction of the angular momentum flow). From
Romanova et al. (2005).}\label{angmom}
\end{figure}
\noindent{\bf Rotational equilibrium state and disk-locking}. If a
magnetized star accretes for a sufficiently long time, it is
expected to be in the {\it rotational equilibrium state}, that is
when the total positive spin acting on the star equals the total
negative spin. Many CTTSs are expected to be in the rotational
equilibrium state (Bouvier 2007). If $R_m=R_{cr}$, then both
inflated and partially-open field lines (which thread the disk at $r
> R_{cr}$) will lead to spin-down of the star. So in the rotational
equilibrium state, $R_{cr}=k R_m$, where $k > 1$.  Numerical
simulations have shown that $k=1.2-1.5$ (Romanova et al. 2002; Long
et al. 2005). However in both cases matter dominates in the corona
and this led to quite large angular momentum outflows associated with
inflated and partially open field lines. In the best cases of Long et
al. (2005) the coronal density is $3\times 10^{-4}$ of that in the
disk. If the corona above the disk has this density (say, due to winds
from the disk or the star) then the estimations done in this paper
are correct.

\section{Modeling of jets and winds from CTTSs}

Jets and winds are observed from a number of CTTSs (e.g Cabrit et
al. 2007). In some cases only slow jets are observed Brittain et al.
2007). Clear outflow signatures are also seen as absorption features
in the blue wings of some lines, such as the He I line which shows
outflow of the order of up to 10\% of the disk mass (Edwards et al.
2003; Kwan et al. 2007; see also Hartmann 1998).

\noindent{\bf Winds from slowly rotating stars}. In early
simulations of slowly rotating stars we did not observe any
significant outflows from the disk or from the disk-magnetosphere
boundary (Romanova et al. 2002, Long et al. 2005). It is clear that
an x-type configuration is favorable for launching outflows from the
disk-magnetosphere boundary (Shu et al. 1994). To obtain this
configuration, we suggested that the accretion rate is initially low
but later increases, and matter comes to the simulation region
bunching field lines into an x-type configuration. In addition we
suggested that the $\alpha-$ parameters regulating viscosity,
$\alpha_v$, and diffusivity, $\alpha_d$, are not very small,
$\alpha_d=0.1$ (compared to earlier research, where we typically
chose $\alpha\approx 0.02$. In addition, we chose the viscosity to
be several times larger than the diffusivity, so that the bunching
of field lines would be supported by the flow. An increase of
$\alpha_d$ from 0.02 to 0.1 helped obtain a reasonable rate of
penetration of incoming matter through field lines. In this case we
obtained a quasi-periodic, x-type wind (see an example of
simulations in \fig{xwind}, where $\alpha_d=0.1$ and
$\alpha_v=0.3$). We observed multiple outbursts of matter into the
wind, driven by a combination of magnetic and centrifugal forces
(Romanova et al. 2008, in preparation). Typical velocities in the
outflow are $v=30-60 km/s$, for typical parameters of CTTS and up to
$20\%$ of the disk matter flows to the wind. The simulations run
long, about $800$ rotations at $r=1$.  Similar conical x-type
outflows have recently been observed with more general initial
conditions. It looks like some outflows are also observed by
Bessolaz et al. (2007) and Zanni et al. (2007). We should note that
in their runs diffusivity is even higher. It seems that relatively
high diffusivity is a necessary condition for getting outflows from
the disk-magnetosphere boundary.

\begin{figure}
\centering
\includegraphics[height=2.1in,angle=0]{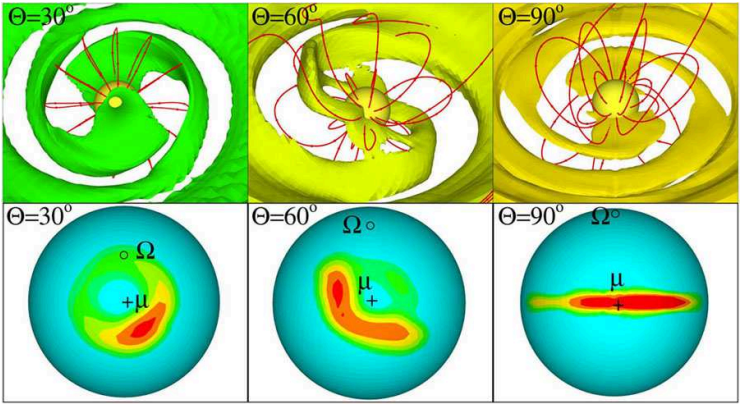}
  \caption{3D simulations of accretion to a star with a misaligned
  dipole magnetic field at misalignment angles $\Theta=30^\circ$,
$\Theta=60^\circ$, $\Theta=90^\circ$. Top panels show matter flow in
the inner part of the simulation region and magnetic field lines.
Bottom panels show corresponding hot spots on the surface of the
star, where the color background shows kinetic energy density. From
Romanova et al. (2004). }\label{3d-fun-spot}
\end{figure}

\noindent{\bf Winds from rapidly rotating stars}. If $R_m > R_{cr}$
then the star is in the ``propeller" regime (e.g., Illarionov \&
Sunyaev 1975; Lovelace, Romanova \& Bisnovatyi-Kogan 1999). This
regime may occur if the accretion rate decreases, causing $R_m$ to
increase. Or, it may be a typical stage during the early years of
the CTTSs evolution. Axisymmetric simulations of the propeller stage
have shown that once again, at low diffusivity $\alpha_{dif}=0.02$,
there are no outflows, while at larger diffusivity, $\alpha_{dif} >
0.1$, conical quasi-periodic or episodic outflows form and carry
away a significant part of the disk matter as wind (see \fig{prop},
also Romanova et al. 2005; Ustyugova et al. 2006). Fig 3 (left
panel) shows that a significant part of the angular momentum carried
by matter is redirected into conical outflows. Fig 3 (right panel)
shows that large part of the angular momentum flows into the corona
along the open magnetic field lines (through the magnetic tower)
 and some flows from the disk to the corona. In the propeller regime we
 were able to obtain a magnetically-dominated corona, and Fig. 3
 shows an example in which the star loses about half of its angular momentum
 through the magnetic wind.
 The propeller stage may be important in fast spin-down of young
 CTTSs.

%%%%%%%%%%%%%%%%%%%%%%%%%%%%%%%%%%%%%%%%%%%%%%%%%%%%%%%%%%%%%%%%%%%%5
\begin{figure}
\begin{center}
\includegraphics[clip,width=0.8\textwidth,angle=0]{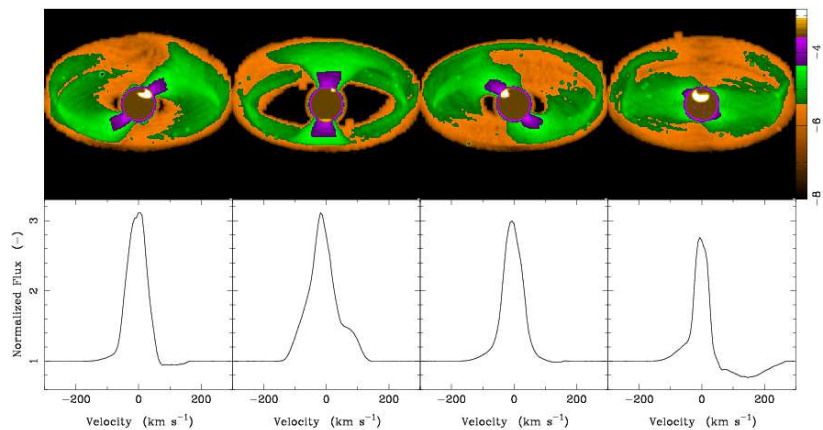}
\end{center}
 \caption{\label{fig:pa-image-line}Pa$\beta$ model intensity maps
  (upper panels) and the
   corresponding profiles (lower panels) computed at rotational phases $t=0.0$, $0.25$,
   $0.5$ and $0.75$ (from left to right) and for inclination $i=60^{\circ}$.
   The misalignment angle of the magnetic axis is fixed at
   $\Theta=15^{\circ}$. The intensity is shown on a logarithmic scale with
   an arbitrary units.  From Kurosawa et al. (2008).}
\end{figure}
%%%%%%%%%%%%%%%%%%%%%%%%%%%%%%%%%%%%%%%%%%%%%%%%%%%%%%%%%%%%%%%%%%%%%%%%%%%%%%%%%%%

%%%%%%%%%%%%%%%%%%%%%%%%%%%%%%%%%%%%%%%%%%%%%%%%%%%%%%%%%%%%%%%%%%%%%%%%%%%%%%%%%%%
\begin{figure}
\begin{center}
\includegraphics[clip,width=0.6\textwidth]{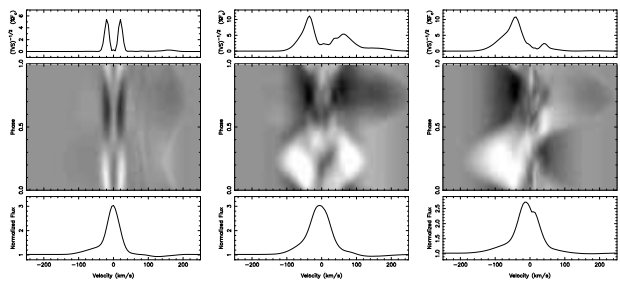}
\end{center}
\caption{\label{fig:variability}The summary of the Pa$\beta$ spectra
computed for the MHD model with $\Theta=15^{\circ}$ and $i =
10^{\circ}, 60^{\circ}$ and $80^{\circ}$ respectively from left to
right.  The spectra were computed at $50$ different rotational
phases. In the bottom panels, the mean spectra of all rotational
phases are shown. In the middle panels, the quotient spectra (each
spectrum divided by the mean spectrum) are shown as grayscale images
with increasing rotational phases in the upward vertical direction.
The temporal variance spectra $(\mathrm{TVS})^{-1/2}$ are shown in
the top panels. From Kurosawa et al. (2008). }
\end{figure}
%%%%%%%%%%%%%%%%%%%%%%%%%%%%%%%%%%%%%%%%%%%%%%%%%%%%%%%%%%%%%%%%%%%%%%%%%%%%%%%%%%%

\section{Properties of the funnel streams, hot spots and the inner disk}

A wide variety of different features were found in our full 3D MHD
simulations of disk accretion to a rotating magnetized star with a
{\it misaligned} dipole magnetic field (Romanova et al. 2003, 2004).
Simulations were done for a variety of misalignment angles from
$\Theta=0^\circ$ to $\Theta=90^\circ$, where $\Theta$ is the angle
between the magnetic axis of the dipole $\mu$ and the angular
velocity of the star $\Omega$ (which was aligned with the angular
velocity of the disk). Below we summarize some interesting features
observed in 3D simulations: ~{\bf (1)} 3D simulations have shown
that the system becomes noticeably non-axisymmetric even at very
small misalignment angles $\Theta \sim 2^\circ - 5^\circ $. Thus,
most magnetized stars are expected to be non-axisymmetric and will
form funnel flows and non-axisymmetric hot spots.~{\bf (2)} Matter
accretes to the star through funnel streams which are not
homogeneous. The density is largest in the interior regions of the
stream, and decreases outwards, so that the appearance  of the
streams depends on the density: at the largest densities they look
like thin streams. At lower densities the stream is wider. The
matter covers the whole magnetosphere at the lowest densities. ~{\bf
(3)} There are usually two streams which flow to the nearest poles.
Some matter flows to the further pole which often leads to a second,
weaker set of streams. In the case of high $\Theta$, one wide stream
often splits into two streams. ~{\bf (4)} The structure of the hot
spots reflects the structure of the funnel streams: the density and
specific kinetic energy are larger in the central regions of spots.
Thus, hot spots are ``hotter" in the center, and cooler outside.
They have different shapes, ranging from bow-shaped for small
misalignment angles to bar-shaped for large misalignment angles (see
\fig{3d-fun-spot}). ~{\bf (5)} Variability curves from the hot spots
obtained from numerical simulations  for different misalignment
angles $\Theta$ and different inclination angles $i$ of the disk
show a variety of shapes, which may be used to constrain $\Theta$
and $i$. ~{\bf (6)} The density distribution in the disk around the
magnetized star is inhomogeneous. It has always a shape of spirals
(see \fig{3d-fun-spot}. ~{\bf (7)} The inner regions of the disk are
slightly warped which reflects the tendency of matter to flow in the
magnetic equatorial plane. This warped disk may obscure the light
from the star, and this may lead to quasi-periodic variations of
light (Bouvier et al. 2003).

\begin{figure}
\centering
\includegraphics[height=2.0in,angle=0]{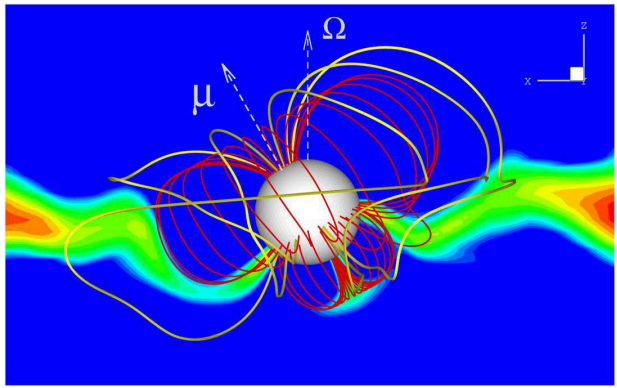}
\caption{3D simulations of accretion to a star with aligned dipole
plus quadrupole fields. Both dipole and quadrupole magnetic moments
are misaligned relative to the rotation axis at $\Theta=30^\circ$.
Here $B_{dipole}\approx 0.45 B_{quad}$. From Long et al. (2007).
}\label{dip-quad}
\end{figure}

\section{Radiative transport and calculation of spectral lines }

We calculated the line profiles from 3D funnel streams using the
radiative transfer code TORUS (Harries 2000; Kurosawa et al. 2004,
2005, 2006; Symington et al. 2005) which was modified to incorporate
the density, velocity and gas temperature structures from the 3D MHD
simulations (Romanova et al. 2003, 2004). The radiative transfer
code uses a three-dimensional (3D) adaptive mesh refinement (AMR)
grid. We take 3D MHD simulations which come to quasi-stationary
state, take parameters of the flow for some moment of time and
calculate spectrum from the funnel streams using TORUS code.

We study the dependence of the observed line variability on two main
parameters: (1) the inclination angle ($i$) and (2) the misalignment
angle ($\Theta$). Here, we show sample results for
$\Theta=15^{\circ}$. As $\Theta$ is rather small,  the accretion
 occurs in two arms, creating two hot spots on
the surface of the star (Romanova et al. 2004; see also
\fig{3d-fun-spot}). As one can see from Fig.~\ref{fig:pa-image-line}
($i=60^{\circ}$)the largest amount of red wing absorption occurs at
the rotational phase at which a hot spot is facing the observer and
when the spot-funnel-observer alignment favorable.

The line variability behavior is summarized in
Fig.~\ref{fig:variability} which shows the phase averaged spectra,
the quotient spectra as a function of rotational phase (in the gray
scale images), and the temporal variance spectrum (TVS), which is
similar to the root-mean-square spectra
(c.f.~\citealt*{fullerton:1996}; Kurosawa et al. 2005).  The mean
spectra of three models are fairly symmetric about the line center;
however, a very weak but noticeable amount of absorption in the red
wings can be seen in the spectra at all $i$.  For $i=10^{\circ}$ and
$60^{\circ}$ cases, the flux levels in their red wing become
slightly below the continuum level, but the level remains above the
continuum for $i=80^{\circ}$ case. Although the line equivalent
width of the mean spectra for $i=10^{\circ}$ is slightly smaller
than that of the $i=60^{\circ}$ and $80^{\circ}$ cases, no major
difference is seen between the three models.

In general, the line profile shapes and strength predicted by the
radiative transfer model based on the MHD simulations are similar to
those seen in the atlas of the observed Pa$\beta$ and Br$\gamma$
given by \citet{folha:2001}. The level of the line variability seen
in the time-series spectroscopic observation (Pa$\beta$) of SU~Aur
(Kurosawa et al. 2005, see their Fig.~1) is comparable to our
models, Fig.~\ref{fig:variability}). The double-peaked variability
pattern (TVS) seen in the low inclination angle model
(i.e.~$i=10^{\circ}$ model in Fig.~\ref{fig:variability}) is very
similar to that of H$\alpha$ and H$\beta$ from the CTTS TW Hydra
observed by \citet{Alencar:2002}. The system has a low inclination
angle ($i=18^{\circ}\pm10^{\circ}$, \citealt{Alencar:2002}) which is
consistent with our model ($i=10^{\circ}$).  Although not shown
here, the line variability predicted from a model with
$i=75^{\circ}$ and $\Theta=60^{\circ}$ resembles that of H$\beta$
from CTTS AA~Tau observed by \citet{Bouvier:2007}. More careful
analysis and tailored model fits of observations are required for
deriving fundamental physical parameters of a particular system.

\begin{figure}
\centering
\includegraphics[height=2.3in,angle=0]{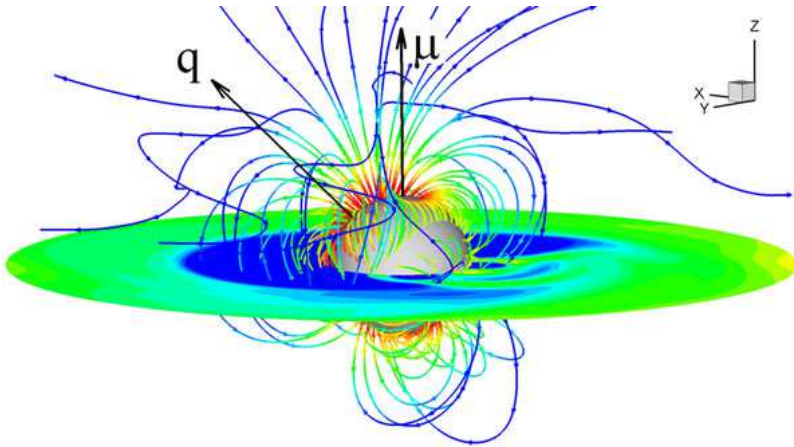}
  \caption{3D simulations of accretion to a star with a complex
  magnetic field: a combination of a dipole
  moment along the z-axis and a quadrupole moment misaligned
  at $\chi=45^\circ$. The equatorial slice shows the density distribution.
  From Long et al. (2008).
}\label{complex}
\end{figure}

\section{Accretion to a star with a complex magnetic field}

The magnetic field of a young T Tauri star may have a complex
structure (see e.g. Johns-Krull et al. 1999; Smirnov et al. 2003;
Gregory, et al. 2006; Jardine et al. 2006). Observational properties
of magnetospheric accretion lead to the conclusion that the magnetic
field of many CTTSs may consist of a combination of dipole and
multipole fields (e.g., Bouvier et al. 2007). If a star has several
magnetic poles, then matter may flow to the star in multiple
streams, choosing the shortest path to the nearby pole. Some
magnetic poles are expected to be closer to the equator of the star
compared with the pure dipole case. The observational properties of
hot spots will also be different. A picture of matter flow to a star
with a complex magnetic field (deduced from Doppler tomography, e.g.
Jardine et al. 2006) has been calculated by Gregory et al. (2006).
It was shown that matter accretes through several funnel streams
choosing, the nearby magnetic poles to accrete.

 Recently
we were able to perform the first 3D MHD simulations of accretion to
a star with a combination of the dipole and quadrupole magnetic
fields (Long, Romanova \& Lovelace 2007, 2008). Simulations of
accretion to a star with a {\it pure quadrupole} field have shown
that most of the matter flows to the star through the quadrupole
``belt", forming a ring-shaped hot spot at the magnetic equator. In
the case of a {\it dipole plus quadrupole} field, the magnetic flux
in the northern hemisphere is larger than that in the southern
hemisphere and the quadrupole belt and the ring are displaced to the
south. \fig{dip-quad} shows the case when the magnetic moment of the
dipole $\rvecmu$ and that of the quadrupole $\mathbf{D}$ are aligned
and both are misaligned relative the the rotational axis
$\mathbf{\Omega}$ at an angle $\Theta$. One can see that the disk is
disrupted by the dipole component of the field but the quadrupole
component is strong enough to guide matter streams to the poles.
 At different
$\Theta$ the light curves have a variety of different features, but
many of these features are observed in the pure dipole cases at
somewhat different $\Theta$ and inclination angle of the system $i$.
So, it would be challenging to deduce the magnetic field
configuration from the light curves (Long et al. 2008).

As the next step we investigated accretion to a star with a dipole
plus quadrupole field, when the magnetic axis of the quadrupole is
misaligned relative to magnetic axis of the dipole at an angle
$\chi$ (Long et al. 2008). \fig{complex} shows an example of such a
accretion for $\chi=45^\circ$.

\begin{figure}
\includegraphics[height=2.3in,angle=0]{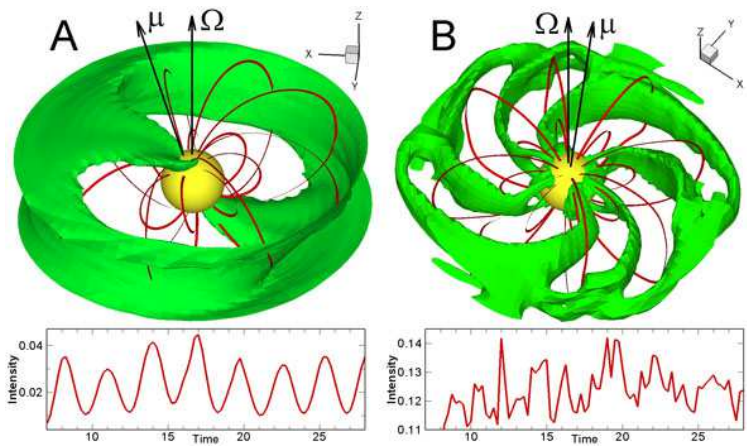}
\centering
  \caption{Left panel: in stable accretion
matter accretes in two ordered funnel streams and the light-curve
has an ordered, almost sinusoidal shape. Right panel: in the
unstable regime matter accretes through multiple tongues and the
light-curve is irregular. From Romanova, Kulkarni \& Lovelace
(2008). }\label{stab-instab}
\end{figure}

\section{Accretion through instabilities}

Recent 3D simulations have shown that in many cases, disk matter
accretes to the star through a 3D Rayleigh-Taylor instability
(Romanova \& Lovelace 2006; Romanova, Kulkarni \& Lovelace 2008;
Kulkarni \& Romanova 2008). Penetration of matter through 3D
instabilities has been predicted theoretically (e.g., Arons \& Lea
1976) and has been observed in 2D simulations by Wang \& Robertson
(1985).  In the unstable regime, matter accretes through the
magnetosphere forming transient but frequently appearing equatorial
tongues (see \fig{stab-instab}, right panel). The tongues penetrate
deep into the magnetosphere and deposit material onto the star much
further away from the magnetic poles than funnel streams do. The
structure of the tongues is opposite to that of the magnetospheric
funnel streams: they are narrow in the longitudinal direction, but
wide in latitude (see \fig{stab-instab}). The number and location of
the tongues varies with time. The number varies between $m=2$ and
$m=7$ and depends on parameters of the model. For example, $m=2$
dominates if the misalignment angle of the dipole is not very small,
$\Theta\sim 15^\circ-30^\circ$. The tongues hit the surface of the
star and form randomly distributed hot spots, so that the light
curve looks irregular. In some cases quasi-periodic oscillations are
observed when a definite number of tongues dominates. Simulations
have shown that this type of accretion occurs in cases when the
accretion rate is relatively high (Romanova et al. 2008; Kulkarni \&
Romanova 2008).

\section{Magnetospheric gaps and survival of protoplanets}

A young star with a strong dipole magnetic field is surrounded by a
low-density {\it magnetospheric gap} where planets may survive
longer than in the disk, due to the low density of matter inside the
gap (if their orbit is inside the 2:1 resonance with the inner
radius of the disk) (Lin et al. 1996; Romanova \& Lovelace 2006; see
sketch at the Figure 10). For typical CTTS the inner edge of the
truncated disk approximately coincides with the peak in distribution
of close-in planets ($P\approx 3 days$). We investigated in 3D
simulations the emptiness of magnetospheric gaps in different
situations.

\noindent{\bf Magnetospheric gaps around stars with  a dipole
magnetic field}. We observed from 3D simulations that in case of the
dipole field and relatively small misalignment angles,  $\Theta <
30^\circ$, the magnetospheric gap has quite a low density (see
Figure 10, right panel). However, for sufficiently large
 $\Theta$, $\Theta > 45^\circ$, part of the funnel stream crosses the
 equatorial plane, so that magnetospheric gap is not empty (Romanova \& Lovelace
 2006).

%%%%%%%%%%%%%%%%%%%%%%%%%%%%%%%%%%%%%%%%%%%

\begin{figure}
\centering
\includegraphics[height=2.3in,angle=0]{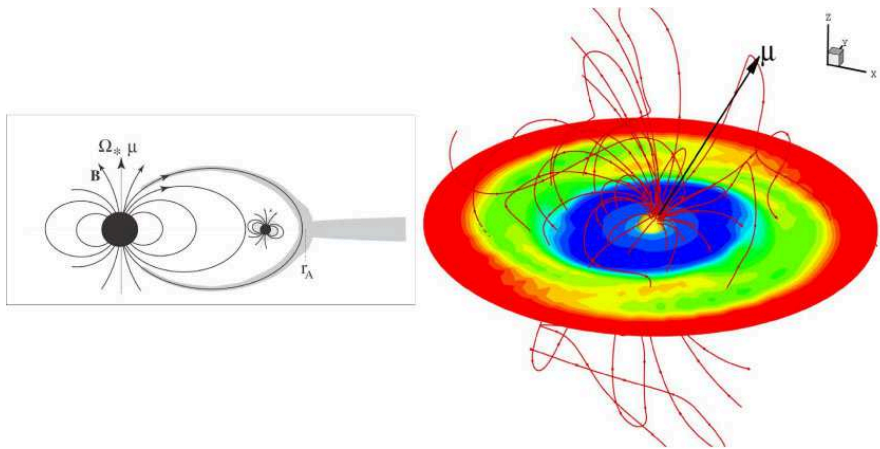}
  \caption{Left panel: sketch of the
magnetospheric gap. Right panel: result of 3D MHD simulations, where
dark blue shows low density, and red - high density with a contrast
of 300. Red lines are sample magnetic field lines. From Romanova \&
Lovelace (2006)}\label{sketch-3d}
\end{figure}

\noindent{\bf Magnetospheric gaps around stars with complex magnetic
fields}. The magnetic field of a young T Tauri star may have a
non-dipole, complex structure (see e.g. Gregory, et al. 2006). If
the dipole component dominates at large distances, and other
components of the field are weaker multipoles, then a magnetospheric
gap may form, as in the case of the dipole field (e.g., Bouvier et
al. 2007). However, if multipoles are sufficiently strong compared
with the dipole field, then matter may accrete through several
streams and will chose a path to the nearby poles which are closest
to the equatorial plane. In this situation some of funnels may cross
the equatorial plane and increase the density in the magnetospheric
gap (see example in \fig{complex}).

\noindent{\bf Accretion through 3D instabilities and magnetospheric
gaps}. At sufficiently high accretion rates in the disk matter
accretes to the star through 3D instabilities (e.g., Romanova et al.
2008; Kulkarni \& Romanova 2008, see \fig{stab-instab}, right panel)
and the magnetospheric gap is not empty. Such accretion is expected
at the early stages of evolution of the CTTSs, when the accretion
rate is expected to be higher. Probably, at this early stage many
protoplanets form, migrate inward, and are absorbed by the star.

\begin{acknowledgments}
We would like to thank organizers for excellent meeting. We also
acknowledge the useful discussion of the disk-locking problem with
Frank Shu, Sean Matt, Jerome Bouvier, and other participants. This
research was partially supported by the NSF grants AST-0507760 and
AST-0607135, and NASA grants NNG05GL49G and NAG5-13060. NASA
provided access to high performance computing facilities.

\end{acknowledgments}


\begin{thebibliography}{}


\bibitem[{Alencar \& Batalha(2002)}]{Alencar:2002}
Alencar, S.~H.~P. \& Batalha, C. 2002, \textit{ApJ}, 571, 378

\bibitem[Aly \& Kuijpers (1990)]{Aly90}
{Aly, J.J., \& Kuijpers, J.} 1990,
 \textit{A\&A}, 227, 473

\bibitem[]{Arons76}
{Arons, J. \& Lea, S. M.} 1976,
 \textit{ApJ}, 207, 914

\bibitem[]{Bessolaz07}
 {Bessolaz, N., Zanni, C., Ferreira, J., Keppens, R., Bouvier, J.,
 Dougados, C.} 2007, \textit{These Proceedings}

 \bibitem[]{Bessolaz08}
 {Bessolaz, N., Zanni, C., Ferreira, J., Keppens, R., Bouvier, J.,
 Dougados, C.} 2008, A\&A, 478, 155

\bibitem[]{Bouvier03}
     {Bouvier, J., Grankin, K. N., Alencar, S. H. P., Dougados, C.,
Fern�dez, M., Basri, G., et al. } 2003,
   \textit{A\&A}, 409, 169

\bibitem[Bouvier, Alencar, Harries, et al. (2007)]{Bouvier07}
     {Bouvier, J., Alencar, S.H.P., Harries, T.J., Johns-Krull, C.M.,
     Romanova, M.M.} 2007,
     in:  B. Reipurth, D. Jewitt, and K. Keil  (eds.),
     \textit{Protostars and Planets V},
     University of Arizona Press, Tucson, vol.\ 951, p.\ 479

\bibitem[{{Bouvier} {et~al.}(2007){Bouvier}, {Alencar}, {Boutelier},
  {Dougados}, {Balog}, {Grankin}, {Hodgkin}, {Ibrahimov}, {Kun}, {Magakian}, \&
  {Pinte}}]{Bouvier:2007}
{Bouvier}, J., {Alencar}, S.~H.~P., {Boutelier}, T., {Dougados}, C.,
{Balog},
  Z., {Grankin}, K., {Hodgkin}, S.~T., {Ibrahimov}, M.~A., {Kun}, M.,
  {Magakian}, T.~Y., \& {Pinte}, C. 2007, \textit{A\&A}, 463, 1017

\bibitem[Bouvier (2007)]{Bouvier07}
     {Bouvier, J.} 2007,
     \textit{These Proceedings}

\bibitem[]{Brittain2007}{Brittain, S., Rettig, T., Balsara, D.,
Tilley, D., Simon, T., Gibb, E., Hinke, K.} 2007, \textit{These
Proceedings}

\bibitem[]{Cabrit07}{Cabrit, S., Dougados, C., Ferreira, J.,
Garcia , P., Raga, A., Agra-Amboade, V., Lavalley, C., Pesenti, M.}
2007, \textit{These Proceedings}

\bibitem[]{Camenzind90}{Camenzind, M.} 1990,
     \textit{Rev. Mod. Astron.}, 3, 234

\bibitem[Edwards, Fischer, Kwan et al. (2003)]{Edwards03}
      {Edwards, S., Fischer, W., Kwan, J., Hillenbrand, L.,
Dupree, A. K.} 2003, \textit{ApJ Lett.}, 599, L41



\bibitem[]{Fendt2000}
   {Fendt, C., \& Elstner, D.} 2000,
     \textit{A\&A} 363, 208


\bibitem[]{Ferreira2006} {
    Ferreira, J., Dougados, C., Cabrit, S.} 2006,
    \textit{A\&A}, 453, 785

\bibitem[{{Folha} \& {Emerson}(2001)}]{folha:2001}
{Folha}, D.~F.~M. \& {Emerson}, J.~P. 2001, \textit{\aap}, 365, 90

\bibitem[{Fullerton {et~al.}(1996)Fullerton, Gies, \& Bolton}]{fullerton:1996}
Fullerton, A.~W., Gies, D.~R., \& Bolton, C.~T. 1996,
\textit{\apjs}, 103, 475

\bibitem[Ghosh \& Lamb (1978)]{Ghosh1978}
     {Ghosh, P. \& Lamb, F. K.} 1978,
     \textit{ApJ Letters} 223, L83

%\bibitem[Ghosh \& Lamb (1979)]{Ghosh1979}
%   {Ghosh, P. \& Lamb, F. K.} 1979,
%     \textit{ApJ} 234, 296



\bibitem[]{Goodson1997}
     {Goodson, A.P., Winglee, R., \& B{\"o}hm, K.H.} 1997,
     \textit{ApJ} 489, 199



\bibitem[]{Gregory06}
    {Gregory, S. G., Jardine, M., Simpson, I., Donati, J.-F.} 2006,
    \textit{MNRAS}, 371, 999

\bibitem[{Harries(2000)}]{harries:2000}
Harries, T.~J. 2000, \textit{MNRAS}, 315, 722


\bibitem[]{Hayashi96}
     {Hayashi, M.R., Shibata, K., \& Matsumoto, R.} 1996,
     \textit{ApJ Letters} 468, L37


\bibitem[]{Hartmann98}
     {Hartmann, L.} 1998,
     \textit{Cambridge University Press}


\bibitem[]{Illarionov1975}
     {Illarionov, A.F., \& Sunyaev, R.A.} 1975,
     \textit{A\&A} 39, 185

\bibitem[]{Jardine06} {Jardine, M., Cameron, A.C., Donati, J.-F.,
Gregory, S.G., Wood, K.}, 2006, \textit{MNRAS}, 367, 917

\bibitem[]{Johns-Krull99} {Johns-Krull, C.M., Valenti, J.A. \& Koresko, C.}
1999, \textit{ApJ} 516, 900

\bibitem[Koldoba, Romanova, Ustyugova, et al. (2002)]{Koldoba 02}
     {Koldoba, A.V., Romanova, M.M., Ustyugova, G.V., Lovelace, R.V.E.} 2002,
     \textit{ApJ Letters} 576, L53

\bibitem[]{Konigl91}
     {Konigl, A.} 1991,
     \textit{ApJ} 370, L39

\bibitem[Kulkarni \& Romanova (2005)]{Kulkarni2005}
     {Kulkarni, A.K. \& Romanova, M.M.} 2005,
     \textit{ApJ} 633, 349


\bibitem[Kulkarni \& Romanova (2008)]{Kulkarni2008}
     {Kulkarni, A.K. \& Romanova, M.M.} 2008, MNRAS
     \textit{accepted} (arXiv:0802.1759)


\bibitem[{{Kurosawa} {et~al.}(2004){Kurosawa}, {Harries}, {Bate}, \&
  {Symington}}]{kurosawa:2004}
{Kurosawa}, R., {Harries}, T.~J., {Bate}, M.~R., \& {Symington},
N.~H. 2004,
  \textit{MNRAS}, 351, 1134

\bibitem[Kurosawa, Harries, Symington (2005)]{kurosawa: 2005}
{Kurosawa}, R., {Harries}, T.~J., \& {Symington}, N.~H. 2005,
\textit{MNRAS}, 358, 671

\bibitem[{{Kurosawa} {et~al.}(2006){Kurosawa}, {Harries}, \&
  {Symington}}]{kurosawa:2006}
---. 2006, \textit{MNRAS}, 370, 580

\bibitem[]{kurosawa07}
{Kurosawa, R., Romanova, M.M. \& Harries, T. J.} 2008, MNRAS,
\textit{accepted}, (arXiv:0802.0201v1 )

\bibitem[Kwan, Edwards \& Fischer (2007)]{Kwan07}
     {Kwan, J., Edwards, S. \& Fischer, W.} 2007,
     \textit{ApJ} 657, 897



\bibitem[]{Lamzin04}
     {Lamzin, S. A., Kravtsova, A. S., Romanova, M. M.\& Batalha, C.}
     2004,
     \textit{Astron. Lett.} 30, 413

\bibitem[]{Lin96}
     {Lin, D.N.C., Bodenheimer, P., \& Richardson, D.C.} 1996,
     \textit{Nature} 380, 606

\bibitem[]{Long 05}
     {Long, M., Romanova, M.M. \& Lovelace, R.V.E.} 2005,
     \textit{ApJ} 634, 1214

\bibitem[Long, Romanova \& Lovelace (2007)]{Long 07}
     {Long, M., Romanova, M.M. \& Lovelace, R.V.E.} 2007a,
     \textit{MNRAS} 374, 436

\bibitem[Long, Romanova \& Lovelace (2008)]{Long 08}
     {Long, M., Romanova, M.M. \& Lovelace, R.V.E.} 2008,   \textit{MNRAS},
     accepted
     (arXiv:0802.2308)

\bibitem[Lovelace, Romanova, Bisnovatyi-Kogan (1995)]{Lovelace 1995}
     {Lovelace, R.V.E., Romanova, M.M., Bisnovatyi-Kogan, G.S.} 1995,
     \textit{MNRAS} 374, 436

\bibitem[Lovelace, Romanova, Bisnovatyi-Kogan (1999)]{Lovelace 1999}
     {Lovelace, R.V.E., Romanova, M.M., Bisnovatyi-Kogan, G.S.} 1999,
     \textit{MNRAS} 514, 368


\bibitem[]{Lynden-Bell2003}
     {Lynden-Bell, D.} 2003,
 \textit{MNRAS} 341, 1360

\bibitem[]{Matt2002}
     {Matt, S., Goodson, A.P., Winglee, R.M., \& B\"ohm,
K.-H.} 2002, \textit{ApJ} 574, 232


\bibitem[]{Matt2005}
     {Matt, S.\& Pudritz, R.E.} 2005, \textit{ApJ Letters} 632, L135

\bibitem[Miller, Stone (1997)]{Miller 97}
     {Miller, K.A. \& Stone, J.M.} 1997,
     \textit{ApJ} 489, 890

\bibitem[Ostriker \& Shu (1995)]{Ostriker 95}
     {Ostriker, E.C. \& Shu, F.H.} 1995,
     \textit{ApJ} 447, 813

\bibitem[Papaloizou \& Terquem (2006)]{Papaloizou 06}
     {Papaloizou \& Terquem} 2006,
     \textit{Rep. Prog. Phys.} 69, 119


\bibitem[]{Pudritz1986}
     {Pudritz, R.E. \& Norman, C.A. } 1986,
     \textit{ApJ} 301, 571

\bibitem[Romanova \& Lovelace (2006)]{Romanova06}
     {Romanova, M.M. \& Lovelace, R.V.E.} 2006,
     \textit{ApJ Lett.} 645, L73



\bibitem[Romanova, Ustyugova, Koldoba, \etal\ (2003)]{Romanova 03}
     {Romanova, M.M., Ustyugova, G.V., Koldoba, A.V., Wick, J.V. \& Lovelace, R.V.E.} 2003,
     \textit{ApJ} 595, 1009

\bibitem[Romanova, Ustyugova, Koldoba, Lovelace (2002)]{Romanova 02}
     {Romanova, M.M., Ustyugova, G.V., Koldoba, A.V. \& Lovelace, R.V.E.} 2002,
     \textit{ApJ} 578, 420

\bibitem[]{Romanova 04}
     --- 2004,
     \textit{ApJ} 610, 920

\bibitem[]{Romanova 05}
     --- 2005,
     \textit{ApJ Lett.} 635, L165

\bibitem[]{Romanova 07}
     --- 2007,
     \textit{in preparation}

\bibitem[Shu, Najita, Ostriker, \etal\ (1994)]{Shu 94}
     {Shu, F., Najita, J., Ostriker, E., Wilkin, F., Ruden, S. \& Lizano, S} 1994,
     \textit{ApJ} 429, 781

\bibitem[] {Smirnov03}
{Smirnov, D.A., Lamzin, S.A., Fabrika, S.N. \& Valyavin, G.G.} 2003,
\textit{A\&A}, 401, 1057

\bibitem[{{Symington} {et~al.}(2005){Symington}, {Harries}, \&
  {Kurosawa}}]{symington:2005}
{Symington}, N.~H., {Harries}, T.~J., \& {Kurosawa}, R. 2005,
\textit{MNRAS}, 356, 1489

\bibitem[]{Ustyugova 06}
     {Ustyugova, G.V., Koldoba, A.V., Romanova, M.M., Lovelace, R.V.E.} 2006,
     \textit{ApJ} 646, 304


\bibitem[]{Uzdensky2002}
 {Uzdensky D.A.} 2002, \textit{ApJ}, 572, 432

\bibitem[]{Wang85}
{Wang, Y.-M. \& Robertson, J. A.} 1985, \textit{ApJ}, 299, 85

\bibitem[] {Weber1967}
{Weber, E.J. \& Davis, L.J. 1967}, \textit{ApJ}, 148, 217

\bibitem[] {Yelenina06}
{Yelenina, T. G., Ustyugova, G. V., Koldoba, A. V.}, \textit{A\&A}
2006, 458, 679

\bibitem[Zanni, Bessolaz \& Ferreira (2007)]{Zanni07}
{Zanni, C., Bessolaz, N. \& Ferreira, J.} 2007, \textit{These
Proceedings}



\end{thebibliography}
\end{document}